\documentclass[twocolumn,11pt]{article}
\usepackage{epsfig}
\usepackage{natbib}
\begin{document}
\bibliographystyle{mnras}

\title{The low surface brightness extent of the Fornax Cluster}
\author{A. Kambas, J.~I Davies, R.~M Smith, S. Bianchi and J.~A Haynes\\
Department of Physics and Astronomy, University of Wales 
Cardiff} 
\date{}
   \maketitle

\begin{abstract}

We have used a large format CCD camera to survey the nearby Fornax cluster and its immediate environment for low luminosity low surface brightness galaxies. Recent observations indicate that these are the most dark matter dominated galaxies known and so they are likely to be a good tracer of the dark matter in clusters. We have identified large numbers of these galaxies consistent with a steep faint end slope of the luminosity function ($\alpha\approx-2$) down to $M_{B}\approx-12$. These galaxies contribute almost the same amount to the total cluster light as the brighter galaxies and they have a spatial extent that is some four times larger. They satisfy two of the important predictions of N-body hierarchical simulations of structure formation using dark halos. The luminosity (mass ?) function  is steep and the mass distribution is more extended than that defined by the brighter galaxies. We also find a large concentration of low surface brightness galaxies around the nearby galaxy NGC1291.
\end{abstract}

\begin{small}
\end{small}

\section{Introduction}

Over the last few decades new populations of galaxies have revealed themselves through the use of high sensitivity detectors at dark observing sites.  From dwarfs to giants, galaxies of Low Surface Brightness (LSB) are a strikingly new addition to the galactic inventory of the Universe (\citealt{iau99}, \citealt{imp97}). Having identified this new class of object we have now moved on into the era of defining their cosmological importance in relation to the brighter, more well known galaxies. To do this it has been necessary to survey significant areas of sky at LSB levels (e.g. \citealt{mor99a}, \citealt{mor99b}, \citealt{ber95}, \citealt{sch95b}, \citealt{imp96}, \citealt{dal997}, \citealt{kash98}). In this paper we extend this survey work to a study of the nearby Fornax cluster and its immediate environment (see also \citealt{cal83}, \citealt{dav88a}, \citealt{fer90}, \citealt{irwin90}, \citealt{both91}, \citealt{hel94}, \citealt{mor99a}, \citealt{mor99b}).

Two fundamental cosmological issues are the spatial scale over which mass is distributed in the Universe and the mass spectrum of the collapsed objects within the Universe. Given that dark matter dominates the mass the challenge is to measure these quantities for the dark, not visible, matter. One obvious solution is to measure the spatial distribution and mass spectrum of the most dark matter dominated galaxies we know of: these are the LSB and dwarf galaxies (\citealt{mcg998}, \citealt{irw95}, \citealt{car88}, \citealt{dek86}). 

Historically the intrinsically bright and high surface brightness galaxies have been used to define the spatial scale and distribution of structures in the Universe (see for example \citealt{mad90}, \citealt{da98}). Their clustering length scale, as measured by such techniques as the two point spatial correlation function, defines the bright galaxy mass scale. The observed clustering scale is smaller than that inferred from simulations of structure growth using currently popular CDM models of dark haloes (see for example \citealt{fren96}, \citealt{jen98}) and so it is often assumed that the bright galaxies are biased tracers of the underlying mass (\citealt{dek87}, \citealt{fren96}). In these models the bright galaxies are more concentrated than the dark matter because they form at the peaks of the initial density fluctuations. They also undergo more mergers than the objects in the lower density outer regions of clusters and so they can become more 'visible' due to enhanced star formation. In this paper we intend to show that, for one nearby cluster at least, there is a population of very LSB low luminosity galaxies that contribute about the same luminosity to the cluster as the brighter galaxies, but which defines a much more extended spatial scale than the bright galaxies. Our expectation is that these galaxies are dark matter dominated (cf. the "dark galaxy" of \citealt{car88}), having mass-to-light ratios of 100 or more similar to the mass-to-light ratios determined for galaxy clusters (\citealt{dav95}).

It is also difficult to relate the mass spectrum inferred from numerical simulations of structure growth (for example CDM) to what we observe. The simulations model the mass spectrum of the dominant dark matter haloes yet what we observe is the luminosity function of galaxies. These two are related through the total mass-to-light ratio of each object, but the mass-to-light ratio is almost certainly a function of galaxy luminosity, type etc. and so relating luminosity to total mass is not easy. In addition, various forms of evolution (mergers, tidal stripping, star formation, galactic winds etc.) may have altered the mass spectrum over time.  To overcome these problems numerous N-body simulations incorporating ever more complicated physics have been carried out in an attempt to relate the initial conditions (\citealt{prs74}) to what we observe (see for example \citealt{fren96}, \citealt{kauf97}). Even with these complications the models of hierarchical structure formation with galaxy evolution still predict a steep faint end slope to the local mass function, similar to the presumed initial conditions. This appears to be a rather robust conclusion. The local cluster galaxy mass function predicted by the models has a low mass slope of $\approx-2$ , (\citealt{fren96}). This is much steeper than that generally found for the luminosity function of field galaxies (\citealt{lov92}, \citealt{mar94}, but see also \citealt{lov97}). However there are a number of recent determinations of a steep luminosity function for clusters (\citealt{phil998}, \citealt{val97}, \citealt{smth96}, \citealt{ber95}). We will find observationally for the Fornax cluster a steep cluster luminosity function consistent with the expected mass spectrum inferred from N-body simulations and show that the steep slope extends to intrinsically faint magnitudes ($M_{B}\approx-12$).

Our prime objective in this paper is to compare the spatial distribution and total luminosity of three samples of Fornax cluster galaxies selected by their surface brightness. We define a bright galaxy sample typical of that used to measure the two point spatial correlation function and the galaxy luminosity function, a LSB sample typical of recent work to determine the numbers of LSB dwarf galaxies in clusters and a very LSB (VLSB) sample typical of what can now be detected using a large area CCD. The bright and LSB samples have been discussed previously by us (\citealt{dis90}, \citealt{mor99a}b, henceforth MDS). The VLSB sample is new and so we will start by describing these data.

\section{Observation and image processing of the VLSB sample}
The data were obtained using the Curtis Schmidt 96{\it cm} Telescope at the CTIO during the nights of 21-24th of October 1997, using the 2048$\times$2048 CCD with 2.3 $arc$ $sec$ pixels.  The total area observed was 13.8 $sq$ $deg$, forming a strip  $\sim$9 $deg$ long outwards from the cluster center (NGC1399) towards the lenticular barred galaxy NGC1291 (figure \ref{fig:strip}). We chose this particular strip of sky because MDS (the LSB sample, see below) had identified numerous quite large LSB galaxies in this area of sky around NGC1291.

There are 13 individual fields in the strip, each of $1.3^{o}\times1.3^{o}$. Each field consists of four exposures of 1050{\it sec}, median stacked to give good signal to noise, while keeping saturation to a minimum.  The different exposures were "jittered" around the nominal field center to improve flat fielding.

\begin{figure*}
\centerline{
      \psfig{figure=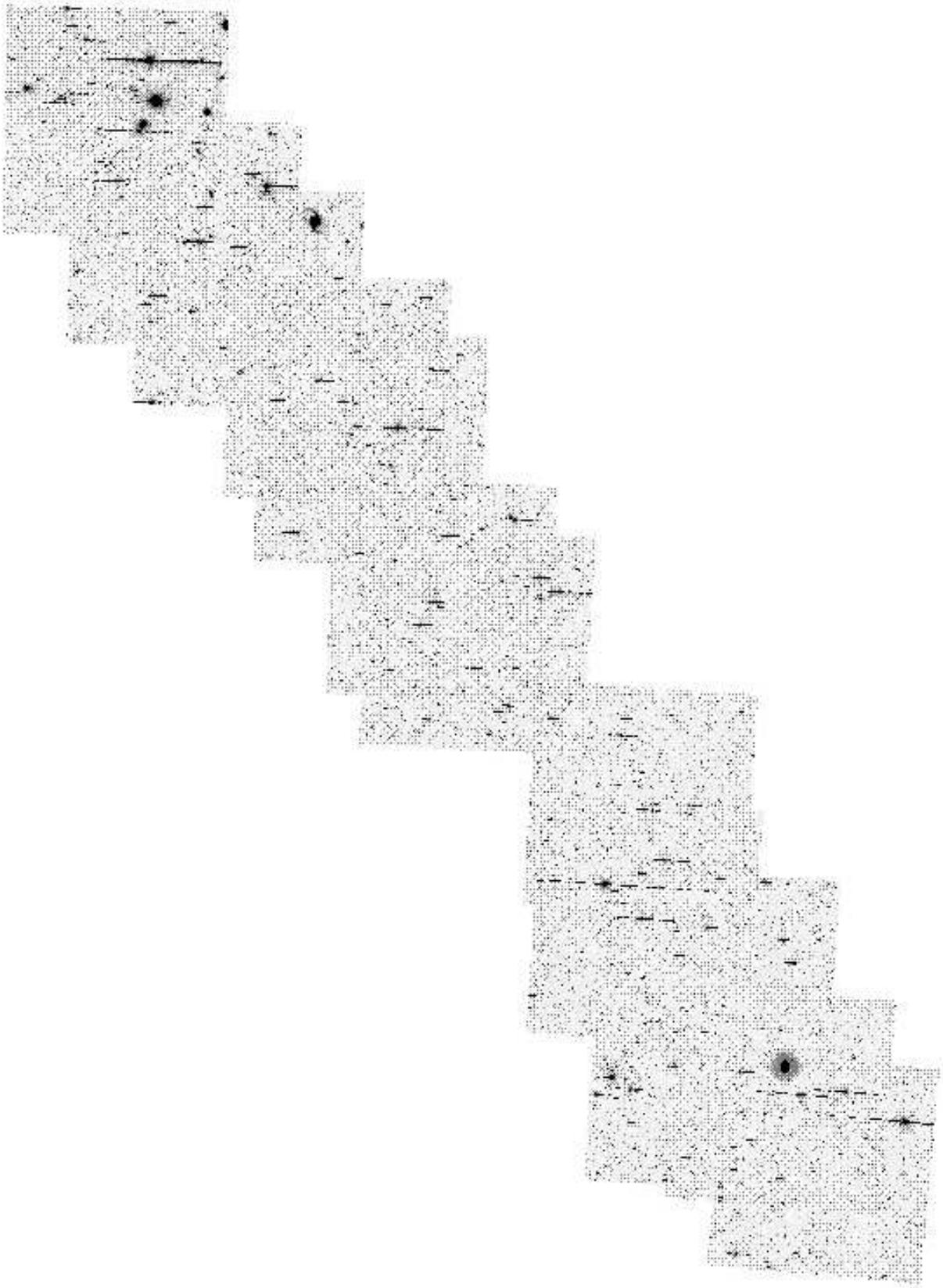,height=22cm}
}
\caption{The strip of sky 9 degrees long from Fornax centre at the top left (NGC1399) out to NGC1291 at the bottom right.}
\label{fig:strip}
\end{figure*}

\subsection{Data reduction}
The CCD uses two read out amplifiers, so all steps of data reduction were performed twice, once for each amplifier. The original images (data, bias frames, flats and standards) were cut into two parts (according to the reading amplifier) and then each side was processed separately. The data remained separated for the whole data reduction and detection process. 

Prior to median stacking, the image data were bias subtracted and flat fielded using the standard techniques. After the stacking and trimming, the size of each field was decreased to $\approx$1 {\it sq deg}. It is very important for the automated detection of LSB and faint galaxies that the sky background is as flat as possible across each frame. Local background variations could lead to the detection of spurious objects in those areas where the sky appears brighter than its global value. In the opposite sense, real LSB objects would not be detected in regions where the local sky is fainter than the overall value (\citealt{jon94}). To overcome this problem we have modelled the sky background on each frame and then subtracted this from the data. The technique consists simply of creating a map of the background sky in which each pixel takes the median value of the pixels in the original frame in an n$\times$n box surrounding the pixel in question. This map is then subtracted from the original data.
The choice of box size is critical, as it will constrain the maximum size of objects that can be detected. Any object with a size approximately half or greater than the box size will be lost or at least severely degraded. The large scale size of the background variations over the frames allowed for a large smoothing box size. A 340$\times$340  pixel box was used. Considering the 2.3  $arc$ $sec$ pixels of the CCD, this translates to objects with angular sizes of about 220 $arc$ $sec$, very much larger than the objects we might expect to detect and our size selection criteria (section 2.2, below). Previous work (\citealt{dav89a}) indicates that Fornax cluster LSB galaxies should have maximum diameters of about 50 $arc$ $sec$. We will show below that all objects detected are much smaller than this upper size limit. This technique allowed us to flatten the sky to $\approx$ 0.5\% which equates to a 1$\sigma$ fluctuation of $\approx$26 $R\mu$.

Calibration was carried out using Landoldt standard stars (\citealt{lan92}) and the standard Harris R filter used for the observations. The observing conditions were not ideal (some high cloud) and we estimate a magnitude error of 0.2, which is sufficiently small for the comparison we intend to make. The sky brightness measured from the frames was $\approx$20.5 R$\mu$, a value typical for the location, season and lunar age. 
Finally, astrometry was carried out using the ESA Guide Star Catalogue and is accurate to about one pixel, 2.3 arc sec. We used this astrometric calibration to measure distances from NGC1399 (the central cluster galaxy) and to obtain surface densities of galaxies in annuli around NGC1399.

\subsection{Image detection and object selection}

The image detection was carried out in two steps. First the frames were scanned for objects using the SExtractor package (Source Extractor, \citealt{bert96}). All objects with 5 connected pixels brighter than 26 $R\mu$ (the brightest 1 sigma isophote from all of the frames) were included in the catalogue.   SExtractor also includes a neural network trained program that classifies each object with a "stellarity index" between 0 (galaxy) and 1 (star). A total of 322,505 objects were detected in this way, giving an average of 23,200 objects per sq deg. The objects were detected by their size with the faintest having an isophotal magnitude of 22.8 $R\mu$. The SExtractor catalogue was then used as the input into our selection program this eliminated any flagged objects, such as ones close to the edges of the frame, near bright stars and composite objects. Also, all objects indexed 1 (definite star) were eliminated at this stage. The catalogue was thus restricted to non-flagged objects with stellarity index up to 0.99 that met the selection criteria. With such large pixels it is difficult to define a star, but the second stage selection described below shows how we finally selected only very extended objects.

\begin{figure*}
\centerline{
\psfig{figure=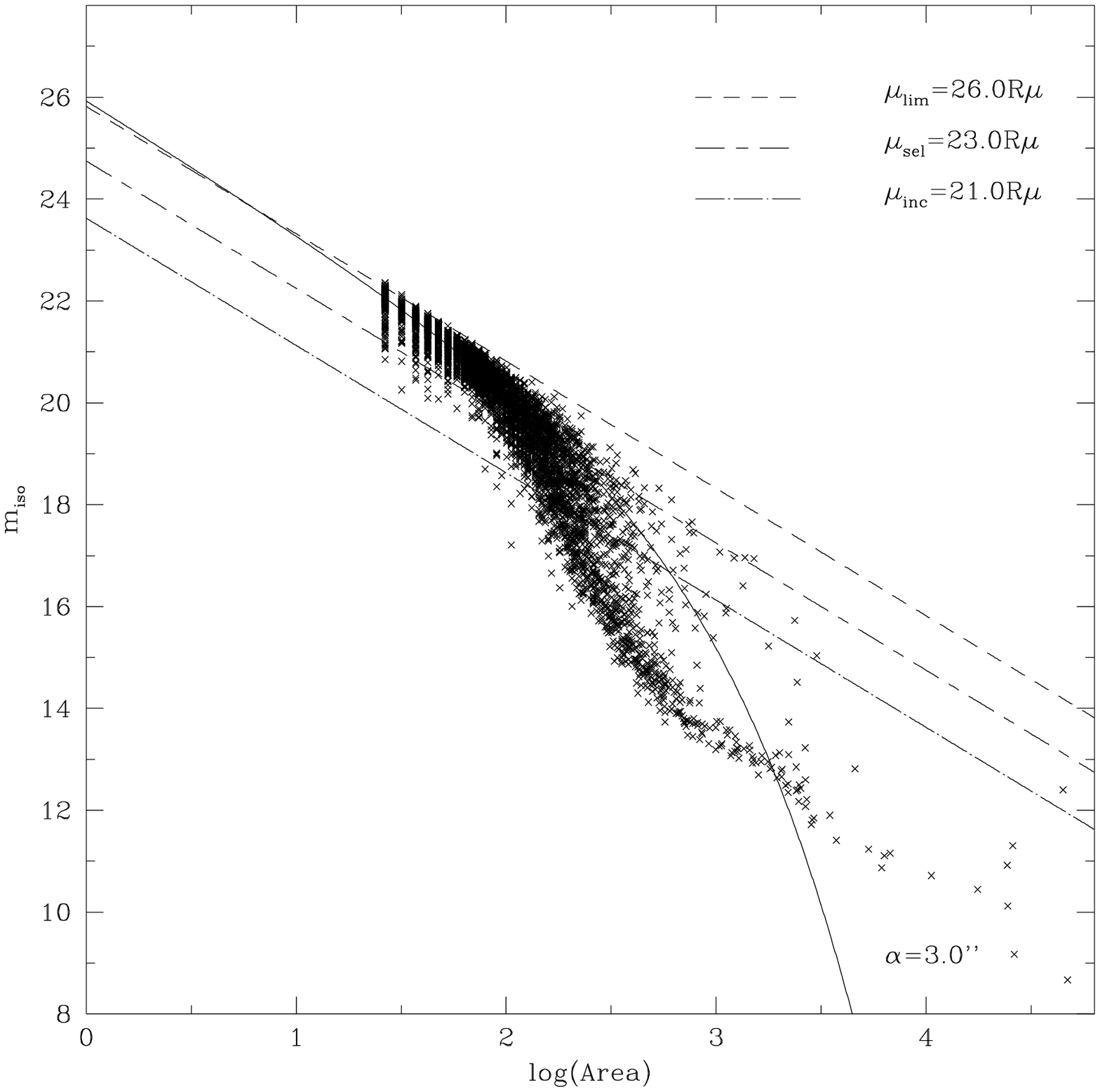,width=6.5cm,bbllx=0pt,bburx=520pt,bblly=80pt,bbury=750pt}
\psfig{figure=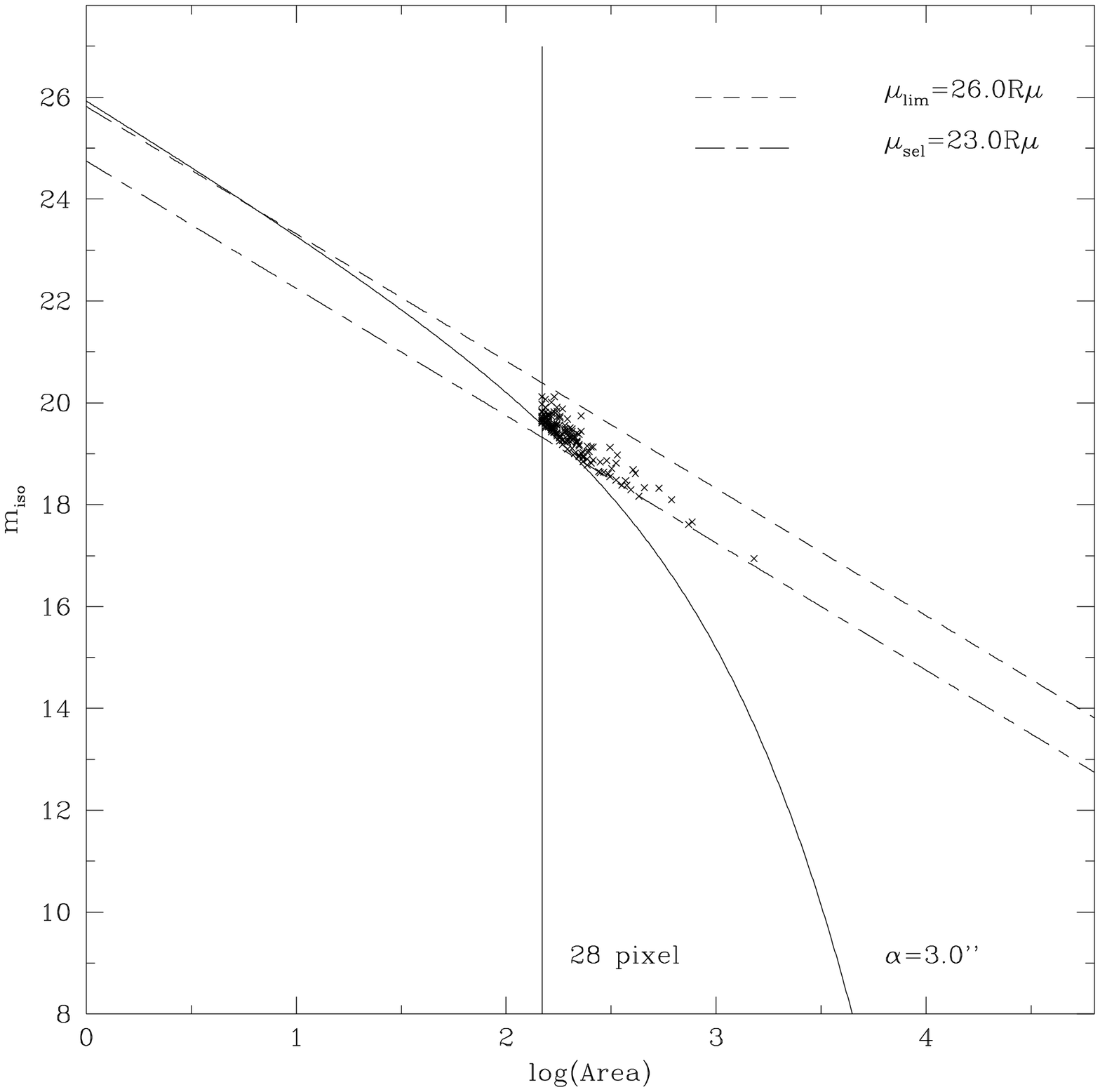,width=6.5cm,bbllx=0pt,bburx=520pt,bblly=80pt,bbury=750pt}
}
\caption{The isophotal magnitude - area plane (log) from one frame. Left: all detected objects. Right: Only selected objects. The two straight lines running diagonally mark the limits in surface brughtness. The lower is our central surface brightness limit of 23 R$\mu$, and the upper is our threshold of 26 R$\mu$. The curve identifies the 3" scalength limit and the vertical line the minimum area criteria.}
\label{fig:sel}
\end{figure*}

The final selection criteria are crucial because we wish to define a sample of galaxies that are predominately nearby cluster members and not distant background galaxies.  Our first criteria was signal-to-noise ratio. To be sure we had real detections we set a minimum signal-to-noise ratio of 10. This equates to a minimum area of $\approx28$ pixels (148 sq arc sec). Using numerical modelling MDS showed that by selecting objects of LSB (in this case $\mu_{o}^{B}>22.5$) and with relatively large exponential scale sizes ($\alpha>3"$) a cluster sample was predominately selected. This conclusion was also reached independently by \citealt{sch96} and \citealt{phil998} (we will return to the question of background contamination in section 3). We also set an upper central surface brightness limit of 23 $R\mu$ so that we have a disjoint surface brightness distribution from the LSB sample (see section 3).  So our final selection criteria for the VLSB sample were:

\begin{enumerate}
\item $\mu_{o}\geq23$ $R\mu$
\item  $\alpha\geq3.0$ arc sec 
\item Isophotal area $\geq148$ sq arc sec
\end{enumerate} 
Plotting the isophotal magnitude of each object against the logarithm of its isophotal area we can draw lines of constant surface brightness and exponential scale size and select only those objects satisfying the above selection criteria (fig. \ref{fig:sel}).

Using these criteria, 3462 objects were selected, an average of 251 objects per sq deg, less than 1$\%$ of the original number of detections.

\subsection{Photometry}

We have used the image parameters produced by SExtractor to determine the photometric properties of each object in the final catalogue. We will assume that the objects detected have exponential light profiles \citep{dav89}.

The mean intensity of a detected object is \\ 
\begin{center}
$<I>=\frac{L_{iso}}{A_{iso}}$\\
\end{center}
where $L_{iso}$ and $A_{iso}$ are the isophotal luminosity and area respectively, \\
\begin{center}
$L_{iso}=2\pi I_{o}\alpha^{2}f$ \\
\end{center}
and \\
\begin{center}
$A_{iso}=\pi r_{L}^{2}$ \\
\end{center}
where the fraction of the total light contained within the limiting isophote is \\ 
\begin{center}
$f=1-e^{-x}(1+x)$ \\
\end{center}
and \\
\begin{center}
$x=\frac{r_{L}}{\alpha}=\frac{\mu_{L}\ -\ \mu_{o}}{1.086}$. \\
\end{center}
$\mu_{L}$ is the limiting surface brightness in magnitudes per sq arc sec and $r_{L}$ is the limiting size at $\mu_{L}$. This leads to the following relationship between the measured mean surface brightness and the exponential central surface brightness. \\ 
\begin{center}
$<\mu>=\mu_{o}-2.5\log f + 5\log x - 0.75$ \\
\end{center}
Thus from the isophotal magnitude and area calculated by SExtractor we can calculate the central surface brightness $\mu_{o}$ (from above), the total magnitude using \\
\begin{center}
$m=m_{iso}+2.5\log{f}$ \\
\end{center}
and the scalelength using
\begin{center}
$\alpha=10^{\frac{\mu_{o}-m-2}{5}}$ \\
\end{center}

It is worth commenting about the influence the large pixels may have had on our determination of the photometric parameters. We do not fit surface brightness profiles. Instead we derive the exponential parameters that correspond to the isophotal luminosity and area measured by SExtractor. Errors would arise if background pixels were included or excluded in the measured area because of the intrinsic fluctuations in the background. The background fluctuations ($\sigma$) are at about the same level as the limiting isophote so, the probability of including or excluding an extra pixel is at most 0.2 for Gaussian fluctuations ($0.2^{n}$ for n pixels). Including or excluding one extra pixel to the area, and ignoring the small contribution to the luminosity leads to an error in the surface brightness of only about 0.05 magnitudes even for the smallest galaxies detected. This will have little effect on our results.

\section{Results}

\subsection{The three samples defined by their surface brightness and magnitude}

In what follows we will compare the clustering scale and the contribution to the total cluster light of three samples selected by their surface brightness. These are three progressively fainter central surface brightness samples. A bright galaxy sample from \citealt{jon80}, a LSB sample from MDS and our VLSB sample describe in this paper. We will also convert the VLSB sample surface brightnesses and magnitudes into the B band using (B-R)=1.5, a value typical of Fornax cluster dE galaxies \citep{eva90}, so that we can  compare this sample with the bright and LSB samples.

\citeauthor{jon80} list 64 Fornax cluster galaxies with m$_{B}\leq16$. We have previously measured the surface brightness distribution of these galaxies \citep{dis90} and we reproduce this in fig. 3. This sample illustrates the familiar Freeman result \citep{fre70}. The distribution of surface brightness is sharply peaked at a value of $\approx$21.5 B$\mu$ with an rms width of $\approx$0.4 B$\mu$. These galaxies will be used to define the luminosity distribution as delineated by the bright galaxies.

The Fornax cluster LSB sample, from the catalogue of MDS, is a small part of a survey covering some 2400 $sq$ $deg$ of sky. The survey was carried out using UK Schmidt telescope photographic plates scanned by the Cambridge (U.K.) Automated Plate Measuring machine. All objectes were objectively selected to satisfy the selection criteria:

\begin{enumerate}
\item {\it $\mu_{o}$}$\geq$22.5 $B\mu$
\item  $\alpha$$\geq$3.0 $arc$ $sec$
\end{enumerate}

This sample is extensively discussed in MDS. The surface brightness distribution is disjoint from that of the bright galaxy sample with a peak at about 23.5 B$\mu$ (fig. 3).

\begin{figure*}
\centerline{
      \psfig{figure=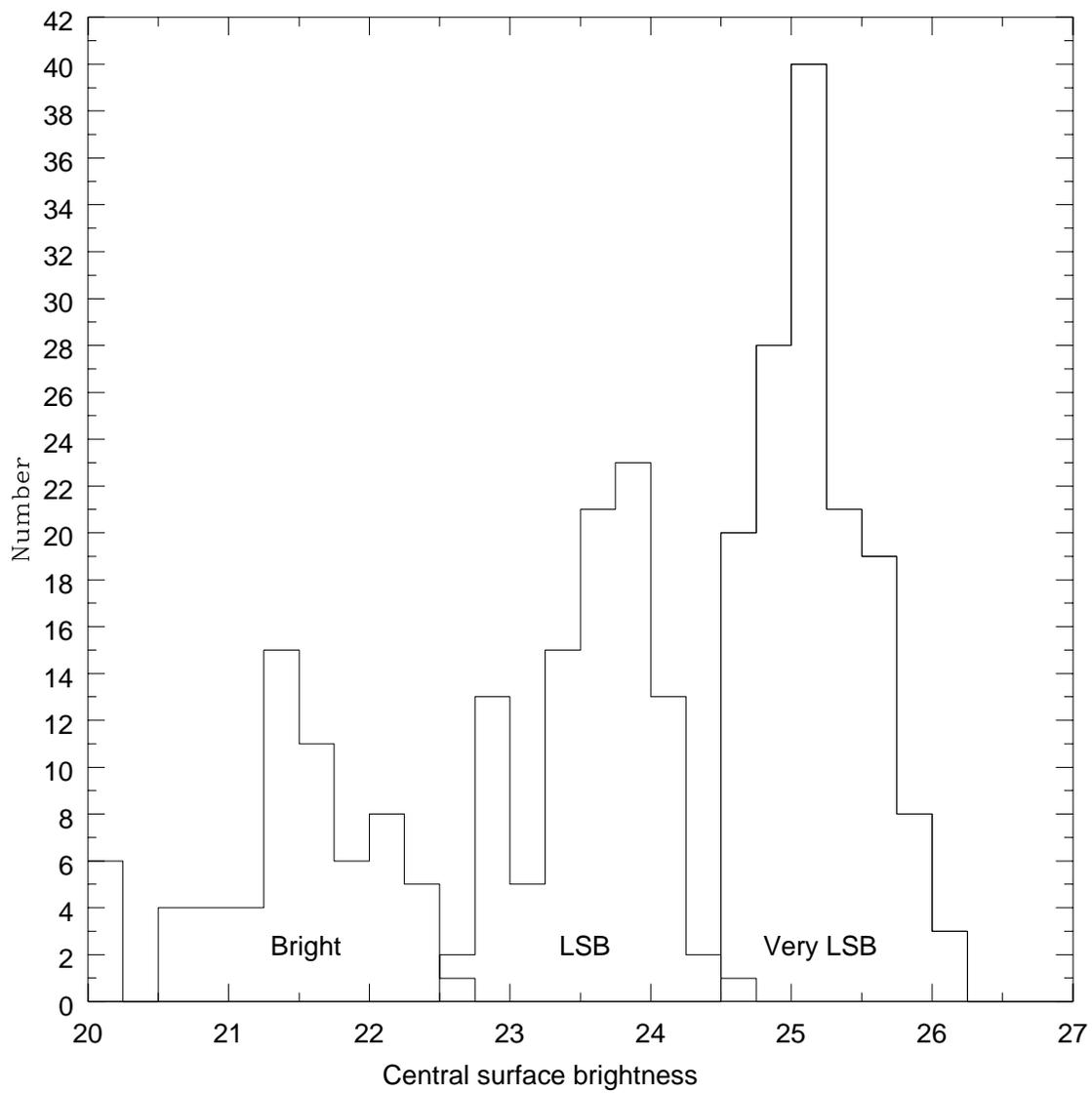,height=16cm}
}
\caption{The B band surface brightness distributions of three different samples of galaxies in the Fornax cluster. The numbers in the VLSB sample have been scaled by a factor of 20 in order to fit on the same plot with the other two samples.}
\label{fig:compsb}
\end{figure*}

The VLSB sample has been discussed in section 2, and its surface brightness distribution is again shown in fig. \ref{fig:compsb}. Note that fig. \ref{fig:compsb} should not be seen as a true distribution of surface brightness for galaxies over some 7 magnitudes. Each sample was selected in a quite different way, at different isophotal levels against different sky backgrounds. Each will be incomplete for galaxies with central surface brightnesses close to the isophotal limit and the higher surface brightness galaxies in each sample will be selected against because at a given magnitude they will be smaller and so they will not satisfy the size criteria. We have made no attempt to correct the sample for surface brightness incompletness because the true distribution of surface brightness is not known. Comprehensive discussions of how such corrections may be made are given in \citealt{dis987}, \citealt{dav90} and \citealt{mcg95}. We will comment further on incompleteness in section 3.3.

The three samples also show a clear trend in apparent magnitude with the lowest surface brightness galaxies having the faintest magnitudes. The mean apparent magnitudes of the three samples are approximately 14.2, 18.1 and 19.8 respectively ($M_{B}$=-17.1, -13.2 and -11.5 for a distance modulus of 31.3, Ferguson 1990).

\subsection{The radial distribution of the three samples}

\begin{figure}
\centerline{
\psfig{figure=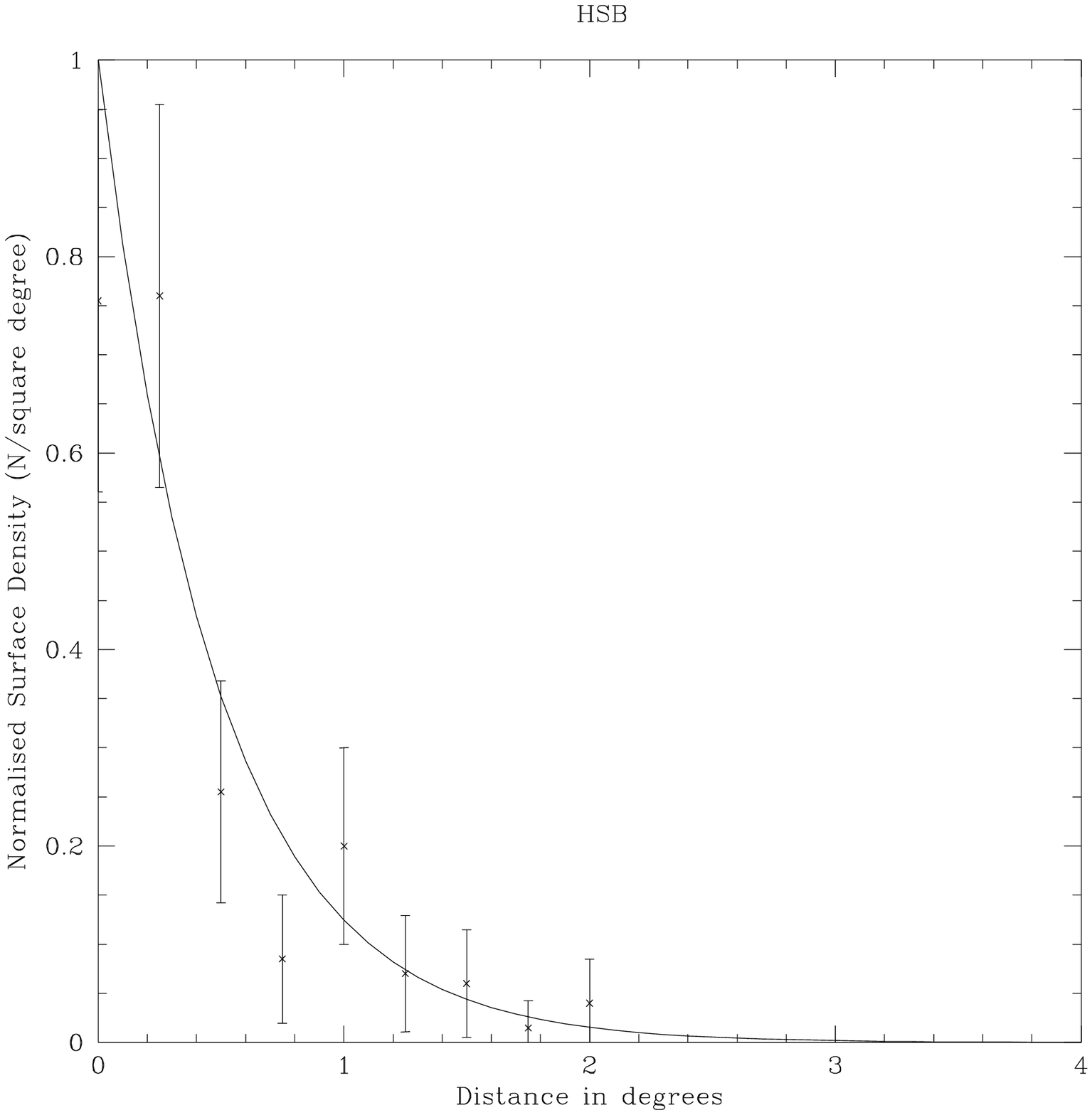,width=5.0cm}}
\centerline{
\psfig{figure=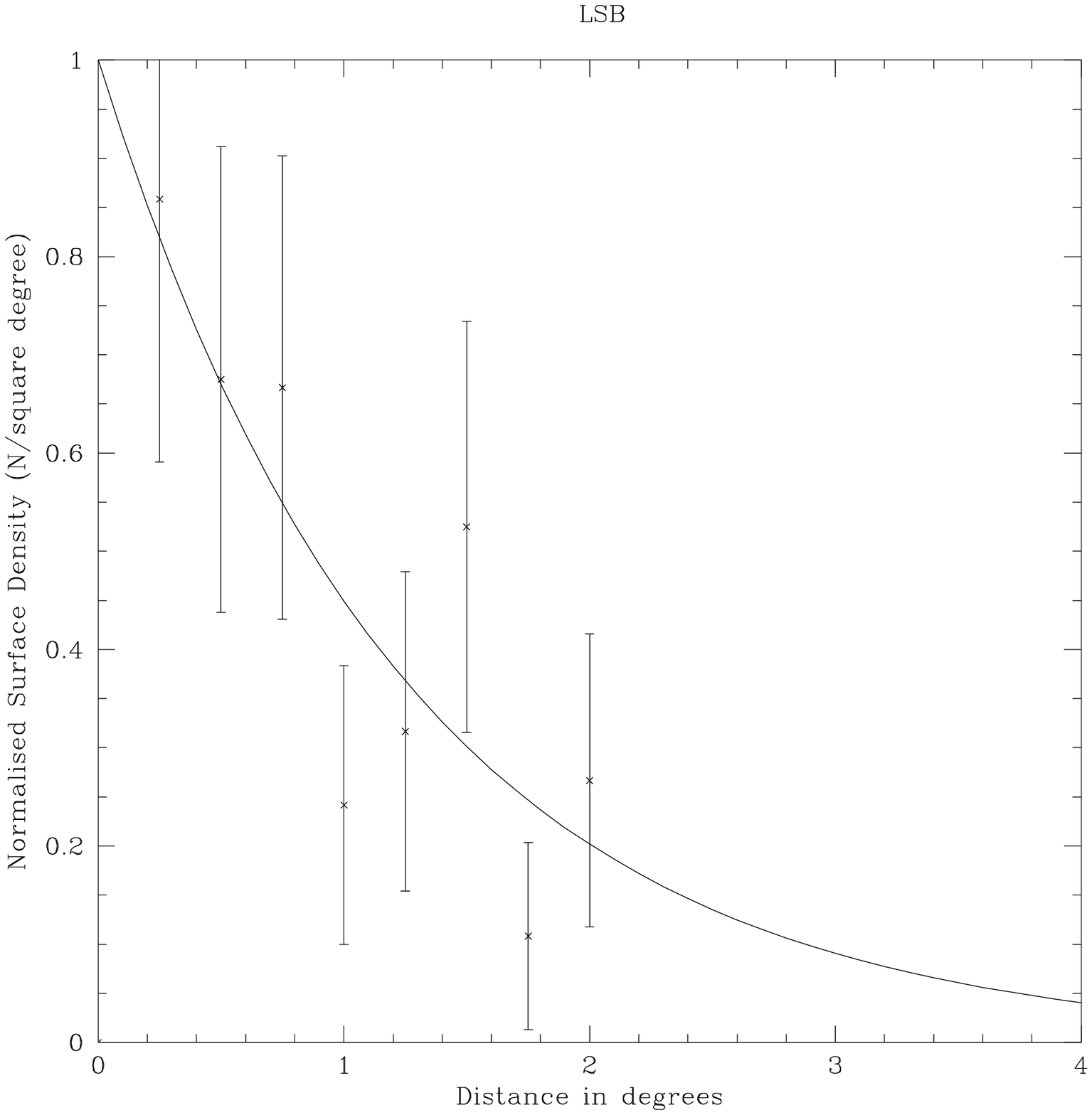,width=5.0cm}}
\centerline{
\psfig{figure=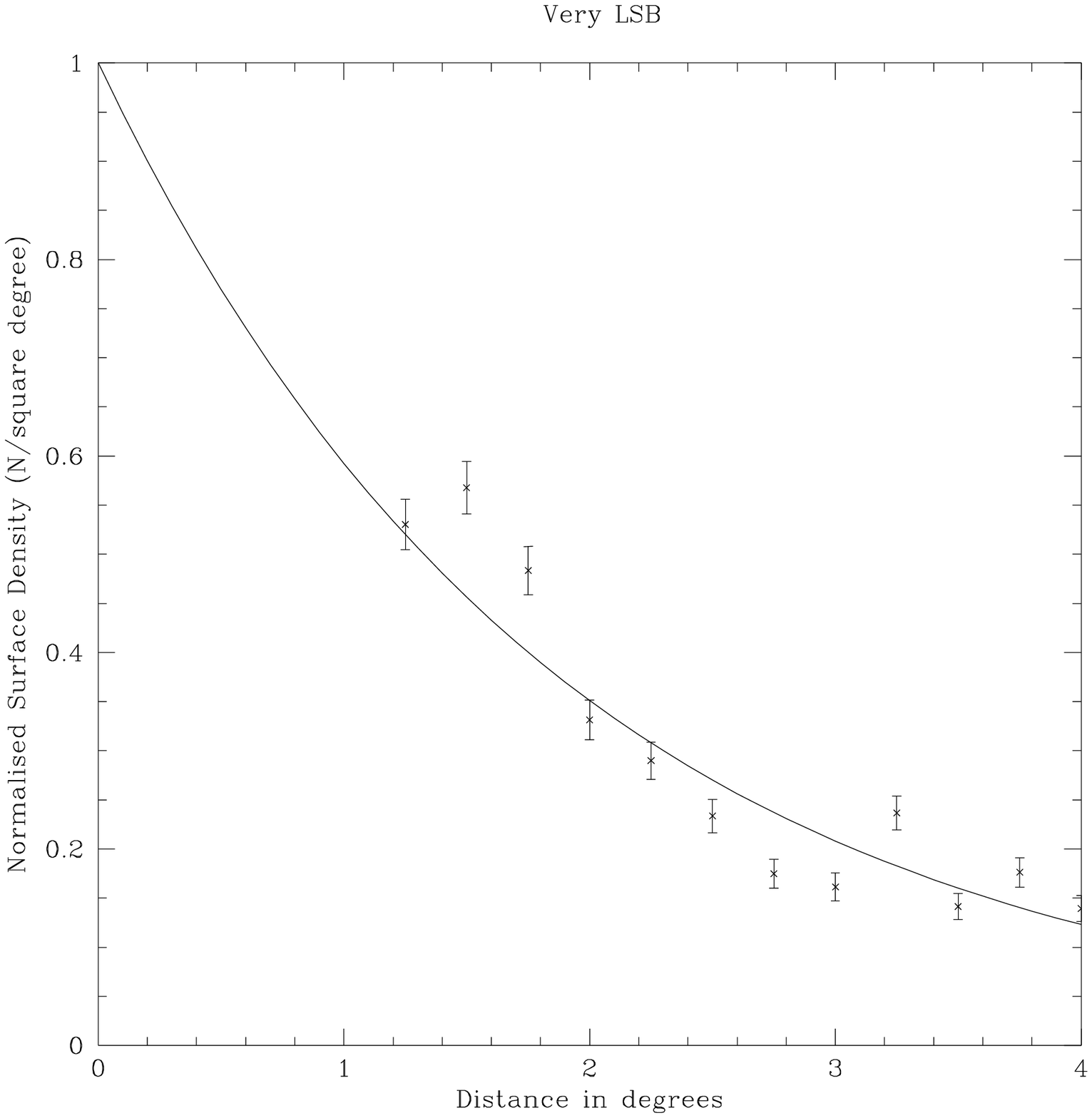,width=5.0cm}}
\centerline{
\psfig{figure=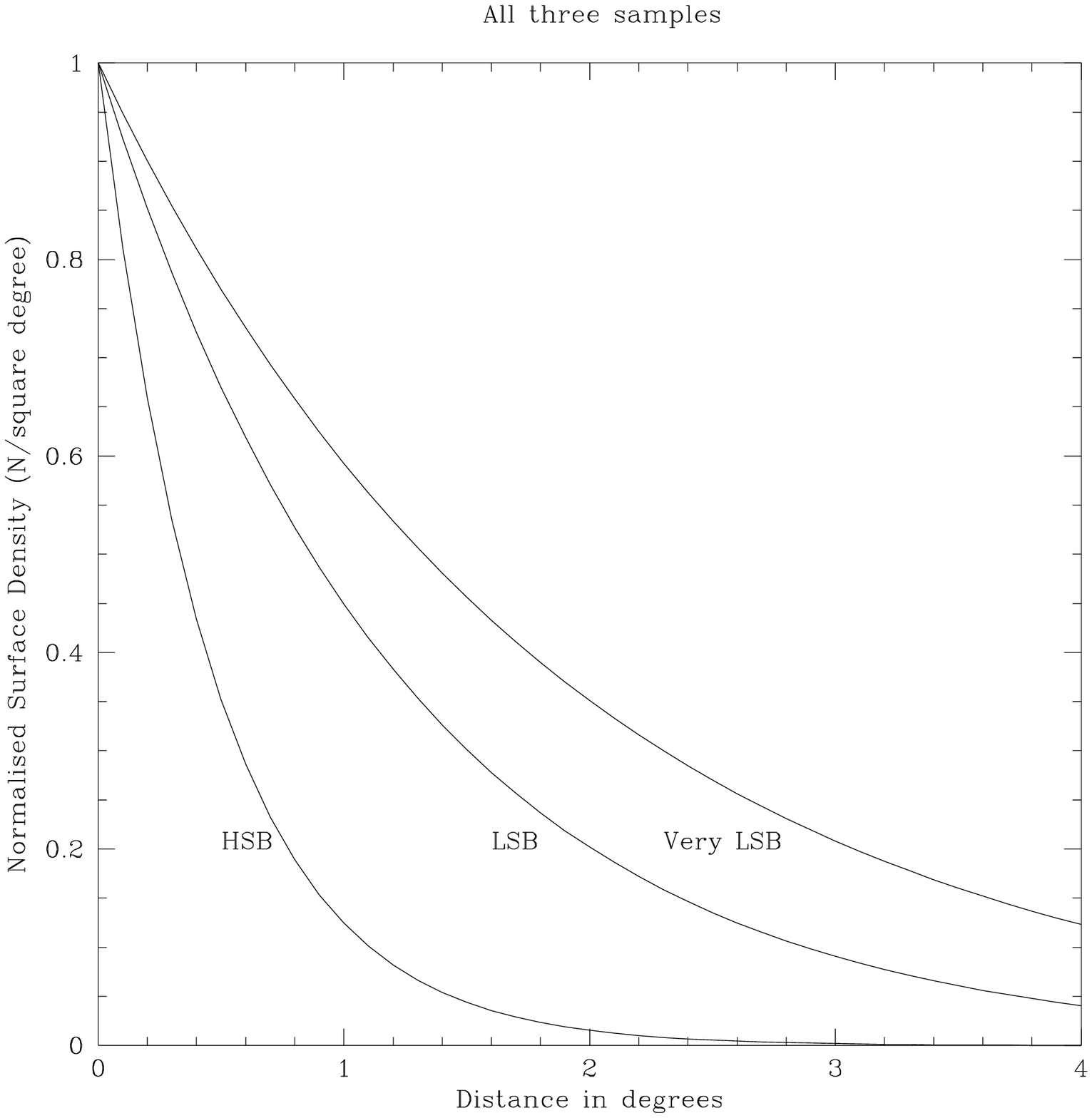,width=5.0cm}}
\caption{The normalised surface density distributions for the three samples. The top three  are the distributions of the three different samples with poissonian errors. The bottom plot is the three samples overlayed, without errorbars for clarity.}
\label{fig:sdd}
\end{figure}

In this and the following section we will assume that all the galaxies in each sample are cluster members. We will justify this assumption at the end of this section. Using the three data sets described above we can measure the decrease in surface number density from the cluster centre, for galaxies of different surface brightnesses (fig. \ref{fig:sdd}). The number density data for the bright and LSB samples are taken from MDS. The data for the VLSB are binned over annulii of 0.25 deg, with all radial distances measured from the central Fornax galaxy NGC1399. Fitting an exponential to the data we find scale sizes of $0.5\pm0.1$, $1.3\pm0.4$ and $1.9\pm0.3$ $deg$ for the bright, LSB and VLSB samples respectively (The central surface density of the VLSB sample is flat within the inner 1 deg and so we do not fit over this region. Including these points would give an even longer scale length). These correspond to approximately 0.16, 0.42 and  0.64  $Mpc$ for a Fornax distance of 18.2 $Mpc$ (Ferguson 1990).  The LSB and VLSB samples are almost 3 and 4 times respectively more spatially extended than the bright galaxy sample. The result for the LSB sample (clustering scales some three times larger than the bright galaxies) has also been shown to be true using a much larger sample of galaxies, over a larger area of sky that includes other nearby galaxy groups (see MDS). It is also consistent with previous work on the spatial extent of more luminous LSB galaxies - they are less strongly clustered than the brighter galaxies (\citealt{mo94}). The exponential fit central galaxy number densities are approximately 40 (HSB), 10 (LSB) and 800 (VLSB) per sq deg. This implies a sharp rise in the luminosity function at the faint end (slope of $\alpha\approx-2$, see below).

\subsection{The contribution to the total luminosity of the cluster}

Having derived the parameters of the exponential fit to the number density distributions we can integrate to determine the total number of cluster galaxies expected in each sample. This corresponds approximately to the total numbers we actually have in the bright and LSB samples. For the VLSB sample the data only extend to large enough distances on one side of the cluster and so the integral takes account of the area not included. We can then use the mean luminosity of each sample to make a crude estimate of the ratio of total luminosities produced at each surface brightness. These are 1.0:0.1:0.5 for the bright:LSB:VLSB samples respectively.  
Adjusting the integral to account for the flattening of the VLSB number density within the central 1 deg makes little difference to the total luminosity (0.5 of the total luminosity becomes 0.45). Integrating to $4^{o}$ rather than infinity reduces the value from 0.5 to 0.3 of the total. Subtracting a background of 100 galaxies per sq deg (see below) alters the 0.5 to 0.3. If the cluster population ends at a radial distance of $\approx 4^{o}$ and there is a background contamination of about 100 galaxies per sq deg then the VLSB sample contributes about 0.1 to the total cluster luminosity. Over the inner $2^{o}$ radius of the cluster the total luminosity ratio  is 1.0:0.05:0.2 ignoring any background contribution. We will argue below that effectively all of the VLSB galaxies belong to the cluster and so we conclude that the VLSB galaxy population contributes a significant fraction of the total cluster light.

It was our initial intention to use the range of luminosities in the VLSB sample to define the faint end of the luminosity function, but it is not possible to do this because of the way the sample has been selected, which itself was a consequence of trying to keep the background contamination to a minimum. The combination of the surface brightness and scale size limit forces a surface brightness size relation upon the data (see fig. \ref{fig:sel}). At a given size we preferentially select faint galaxies because these are the ones that have low enough surface brightness. This would have had the effect of making the luminosity function appear steeper than it really is (this will affect other samples selected in the same way, \citealt{sch96}). We cannot correct for this because we do not know the number of cluster galaxies that satisfy the individual surface brightness limits, but fail to be selected because of their size. For example \citealt{drin99} have identified (using redshifts) Fornax cluster galaxies with approximately the same surface brightness as the galaxies in the bright sample, but with much smaller sizes and fainter magnitudes. These were not included in the \citealt{jon80} magnitude limited sample.

\begin{figure*}
\centerline{
      \psfig{figure=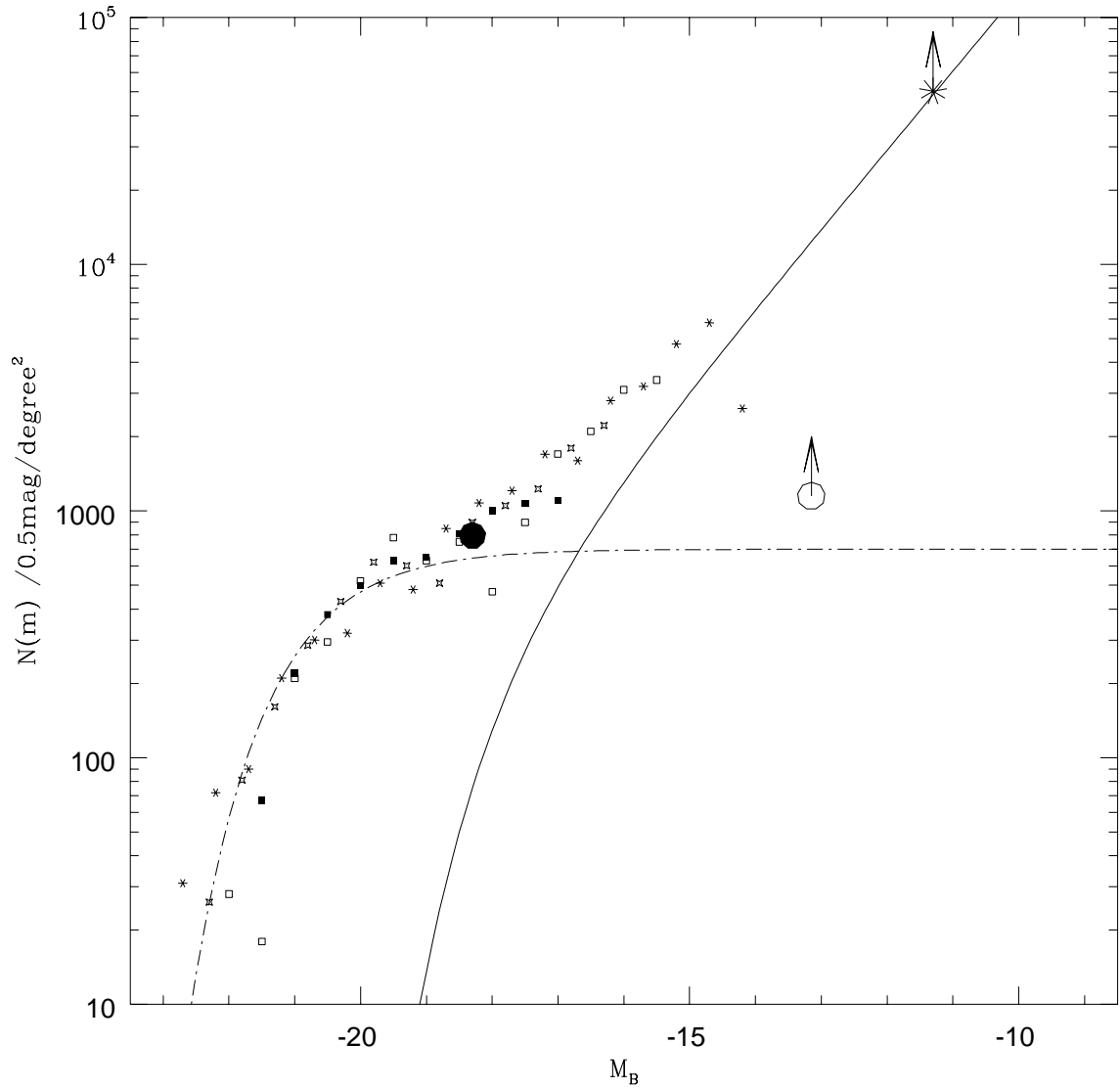,height=16cm}
}
\caption{A comparison of our data  with the luminosity functions derived for other clusters by Smith et al. 1996. The filled circle indicates the bright galaxy sample. The LSB (open circle) and VLSB (star) samples are indicated as lower limits. Two Schecter LF are plotted (the fit to the Smith et al. data), with $\alpha=-1.0$ (dotted) and  with $\alpha=-1.8$ (solid).}
\label{fig:lfcomp}
\end{figure*}

What we have done is to estimate a lower limit to the contribution made by the LSB and VLSB samples to the luminosity function. We have done this by simply using the mean magnitude (standard deviation is about 0.8 mag) and total numbers (integrated to infinity) of each sample. We have then normalised this to the total numbers of bright galaxies. In fig. \ref{fig:lfcomp} we have scaled our numbers so that the bright galaxy numbers correspond to the normalised luminosity functions given in \citealt{smth96} (converted to the B band).  \citealt{smth96} show that the luminosity function of three more distant clusters steepens beyond $M_{B}\approx-19.5$ with a faint end slope of $\approx-1.8$. Our data (lower limits) is consistent with this steep slope and extends this result by some 3.5 magnitudes to $M_{B}\approx-12$. It is consistent with a similar result obtained by \citealt{phil998} for the Virgo cluster.

\subsection{Contamination by non-cluster galaxies}

The above result depends on there being little contamination of our sample by non-cluster galaxies. The radial surface density plots of course indicate a substantial number of cluster galaxies, but there is still the possibility of contamination of the sample by non-cluster galaxies.

The bright sample is one where cluster membership has been assigned because of a measured redshift and so there is no  ambiguity. Possible background contamination of the LSB sample has been extensively discussed in MDS. The conclusion was that at most 20\% of the galaxies could be in the background. 

To try and assess the background contamination of the VLSB sample we have analysed the surface density of galaxies over the full $9^{o}$ extent of the data. The data terminate with one field centred on the peculiar lenticular galaxy NGC1291 (fig.\ref{fig:strip}). Our original motivation for choosing this strip was that MDS had identified a number of relatively large LSB galaxies in the vicinity of NGC1291.
A visual inspection of the field around NGC1291 on the CCD also clearly indicates a large number of LSB galaxies. 

\begin{figure*}
\centerline{
      \psfig{figure=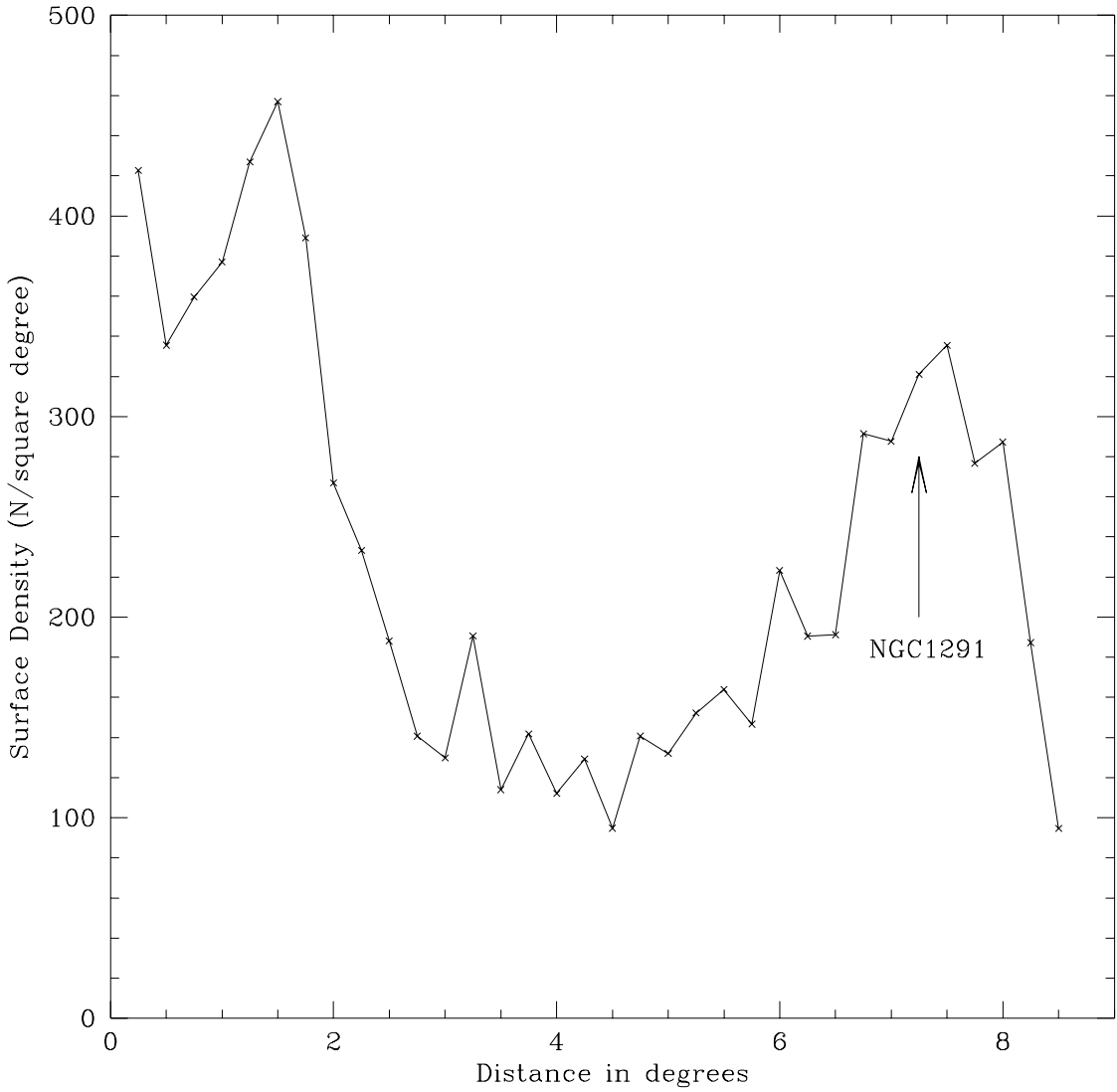,height=16cm}
}
\caption{The surface density distribution extending to 9$^{o}$ from the centre of the cluster.}
\label{fig:dist}
\end{figure*}

 In fig. \ref{fig:dist} we show the normalised surface density of galaxies as a function of distance from the centre of Fornax extending to $\approx9$ $deg$. The most striking feature is the initial decline and then large increase in galaxy numbers as NGC1291 is approached. NGC1291 appears to be at the centre of a cluster of LSB low luminosity galaxies ! One can interpret fig. \ref{fig:dist} in two ways. Either the Fornax cluster population comes to an end at a radius of about 4 deg and there is a background surface density of order 100 galaxies per sq deg or the two extended LSB populations of Fornax and NGC1291 overlap at about 4 deg. 

Numerical simulations of the background contamination of the LSB sample by MDS predict about one background galaxy per sq deg with scale sizes greater than 3 arc sec. The same simulation but with the lower surface brightness cut-off of the VLSB sample predicts zero background galaxies per sq deg (the simulation assumes a flat faint-end to the luminosity function (\citealt{mar94}) and a Gaussian surface brightness distribution with a peak at Freeman's value (\citealt{fre70})). It is easy to show why this is so. A typical L* spiral galaxy would have to be at a redshift of $z\approx0.3$ to have sufficient cosmological dimming to enter the VLSB sample. At z=0.3 its scale size would be less than an arc second. To have an observed scale length of 3 arc sec it would need to be at $z\approx0.07$ at which point it's surface brightness is cosmologically dimmed by only about 0.5 magnitudes. The other option is that the field galaxy luminosity function is similar to that of the cluster ie. that beyond about 4 deg the galaxies in the VLSB sample are no longer associated with the cluster. We believe that there is observational evidence against this idea.

Observations of the general field have been made by \citealt{jon94} and \citealt{dal997}. Both of these surveys were optimised to detect LSB galaxies in the field with similar surface brightnesses and scale sizes to the galaxies in the VLSB sample. The result from  \citealt{jon94} was an upper limit of 10 per sq deg and from  \citealt{dal997} 7 per sq deg. This is far fewer than the 100 per sq deg suggested by our data. So, our conclusion is that these are two overlapping extended haloes of dwarf LSB galaxies about Fornax and NGC1291. We believe that the VLSB sample is essentially a Fornax cluster sample and that our three samples are not adversely affected by background contamination.

\subsection{NGC1291}

NGC1291 is not classified as a Fornax member. It is a peculiar lenticular barred galaxy in the field between us and Fornax at a velocity of 839{\it km $s^{-1}$} \citep{dev75}. The mean velocity of the Fornax cluster is 1366{\it km $s^{-1}$} \citep{fer90}. In the same way as we fitted an exponential to the surface density of galaxies around Fornax we have derived the exponential scale length of the surface density around NGC1291. The value is $\approx$ 1.4 +/- 0.1 $deg$ which equates to 0.22 $Mpc$ at the distance of NGC1291 (The Fornax scale length is 0.64 $Mpc$).  This implies an extended, but not ridiculously large size for a halo of dwarf companions about an isolated galaxy (\citealt{zar93}). We estimate the total luminosity of the VLSB population to be about 0.7 that of NGC1291 (apparent B magnitide of 9.4). This is a marginally larger number than Fornax, but the observations extend some 1.5 magnitudes further down the luminosity function and so we might expect about 4 times more galaxies if the luminosity function carried on with a slope of -2.0, each with about 4 times less luminosity.
This suggests that if you could identify an ultra-LSB population you would conclude that there is more light in the extreme LSB constituents of the cluster than in the bright galaxies. Of course, at some point the luminosity function must turn over or terminate.

It is interesting to calculate the approximate mass contained within the VLSB galaxy population surrounding NGC1291. \citealt{irw95} find that typical (M/L) ratios for dwarf galaxies lie within the range 10-100, whilst that for the central, optical, part of a 'normal' galaxy is in the range 1-10 (e.g. \citealt{ken87}). Given our above  estimate that the total amount of light emitted from the VLSB galaxies is comparable to that from the central galaxy, over ten times as much mass may reside in the VLSB companions. This is comparable to the halo mass of $2\times10^{12}$ $M_{\odot}$ determined by \citealt{zar93} for isolated spiral galaxies. Thus the agreement between the scale-length of the NGC1291 dwarf population and the halo size estimated by \citealt{zar93} together with that between the halo mass and total mass of dwarfs, indicates that the 'dark halo' of some spiral galaxies could be attributed to previously undetected VLSB galaxies.  
 
The VLSB galaxies around NGC1291 may have a primordial origin, but another explanation may be found in the morphology of the bright galaxy; it is a peculiar galaxy. Its morphological classification ranges from "the remarkable barred lenticular galaxy" \citep{dev75} to a ringed SBa (NED) it has a dual personality because it also appears to be known as NGC1269. It is quite possible that NGC1291 has relatively recently undergone a merger event and that some or all the VLSB galaxies are the debris from this collision. 

\section{Discussion} 

The existence of such large numbers of previously undetected galaxies is an important result and ones first thoughts are to other observations that might contradict it. One concern is that we do not seem to find very large numbers of low luminosity VLSB galaxies in the Local Group. The observed luminosity function of the Local Group is relatively flat over the range of magnitudes sampled by our data ($\alpha \approx -1$, \citealt{van992}). Recently \citealt{blitz99} have suggested that the Local Group mass function is really quite steep. They have proposed that the high velocity clouds detected at 21cm are in fact the remains of the smaller dark haloes predicted by the numerical simulations. This remains to be confirmed. It may be that all groups have flat luminosity functions. \citealt{mur98} have compared the luminosity function of galaxies in groups with those in clusters. They found that the luminosity function faint-end slope of the groups was much flatter ($\alpha\approx-1.0$) than that of the clusters ($\alpha\approx-1.4$). So the apparently flat luminosity function of the Local Group compared to a cluster like Fornax may not be unusual or a concern.

 With an observed central galaxy number density of about 450 per sq deg we predict a diffuse intra-cluster light of about 31 $B\mu$ at the centre of the cluster. Such low surface brightness levels would be very difficult to detect against the relatively bright foreground sky (at best 23 $B\mu$). It is unlikely that a large population of low luminosity VLSB galaxies like this could be detected by their integrated light. Inter-cluster stars have previously been identified in Fornax by \citealt{theu97}. These stars may be associated with the cluster low luminosity VLSB galaxies. Considering the selection effects and an assumed  stellar population \citealt{theu97} estimate that these stars could account for 40\% of the total cluster light a similar value to that derived by us for the VLSB galaxy population.

There has been an on-going debate about the nature of the objects that give rise to the absorption features seen in the spectra of quasars (\citealt{lan99}). The issue is whether the Lyman limit and the damped Ly$\alpha$ absorption features arise in normal galaxies (in which case they have very large haloes) or in a previously unseen population of VLSB galaxies that are companions to the brighter galaxies (\citealt{phil93}, \citealt{lin98}). The Fornax VLSB population has a total cross-sectional area on the sky about three times greater than that of the brighter galaxies (equating isophotal sizes). If the VLSB galaxies have sufficiently high gas masses then one would expect the majority of quasar absorption features, seen in sight lines through the cluster, to occur in these small galaxies and not in extended haloes of the brighter galaxies. Recently \citealt{imp99} have compared absorption line features seen in the spectra of quasars behind the Virgo cluster with known cluster galaxies. They point out the ambiguity in trying to assign a given absorption feature with a particular galaxy. They also show that as you select Virgo galaxies of fainter magnitudes you are much more likely to conclude that a low luminosity galaxy is the absorber rather than a bright galaxy with a very large halo. \citealt{imp99} also find that the correlation amplitude of the Ly$_{\alpha}$ absorbers is 4-5 times smaller than that of the brighter galaxies, consistent with the absorption lines occuring in an extended VLSB galaxy population similar to that described here. Observations of Local Group dSph galaxies indicate low gas masses, $\approx10^{7}$ $M_{\odot}$ (\citealt{y97}), but because of the small size of the systems the column densities are not extremely low $10^{19}-10^{20}$ $atoms$ $cm^{-2}$. This is typical of what one might expect through the outer regions of a galactic disc. We conclude Ly$_{\alpha}$ absorption lines could thus arise from an extended distribution of VLSB low luminosity galaxies in clusters or around isolated bright galaxies like NGC1291 rather than in extended gaseous halos.
 
Others have measured steep cluster luminosity functions before (e.g. \citealt{phil998}, \citealt{trent97}, \citealt{wil97}). One concern about these determinations has been the possible contamination by background galaxies. The measured background contamination was statistically removed from the data by assessing the numbers of background galaxies at each magnitude determined from fields away from the cluster. If this had been done incorrectly then a fit to the faint background number counts (slope $\approx0.6$) would lead erroneously to a steep ($\alpha\approx-2.5$) faint-end slope to the luminosity function. Our approach is different. We have specifically selected large LSB galaxies because our models (MDS) indicated very low background contamination if we selected in this way. In addition because of the large extent of our data we have been able to show a decrease in galaxy number density with cluster radius. This strongly supports our contention that we have predominately selected a cluster sample and confirms the previous observations and interpretations of \citealt{phil998}, \citealt{trent97} and \citealt{wil97}.

The clustering scale of the VLSB sample compared to the bright sample is similar to predictions of N-body hierarchical simulations of structure formation with "bias" (\citealt{whi91}, \citealt{fren96}, \citealt{kauf97}). For example many features of the numerical simulations of a galaxy cluster by \citealt{fren96} are seen in the observational data. The galaxies show mass segregation with the most massive galaxies being more centrally clustered in relation to the smaller galaxies and to the dark matter (by about a factor of 3). Also the low mass slope of the simulated mass function is  $\approx-1.8$.
To be the dominant mass component of the cluster the VLSB galaxies require large values of (M/L). Previous work (\citealt{mcg998}, \citealt{car88}) has suggested that LSB and dwarf galaxies are the most dark matter dominated galaxies of all. For example the mass to light ratios of LSB dwarf galaxies in the Local Group can be large ((M/L)$_{\odot}$=10-200, \citealt{irw95}). So, the VLSB galaxy population could account for all of the dark matter in clusters if the galaxy mass to light ratio is similar to the cluster mass to light ratio (of order 100-150, \citealt{dav95}) or there are significantly more galaxies below our present detection limits. The VLSB galaxies may well be the most reliable tracers of the dark matter content of clusters and so we might expect the dominant dark matter of a cluster to reside in its difficult-to-detect very LSB, low luminosity galaxy population.

\section{Conclusions}
We have identified a faint VLSB galaxy population associated with the nearby Fornax cluster. 
The VLSB galaxies appear to be associated with the brighter cluster galaxies, but their distribution is more spatially extended. 
Integrating the total numbers we estimate that they contribute almost as much to the total luminosity of the cluster as the brighter galaxies and they have a larger cross-sectional area on the sky. If they have mass-to-light ratios as high as some more nearby dwarf galaxies they will dominate the mass of the cluster.
Faint galaxies like these have previously been predicted by extrapolations of recently determined cluster luminosity functions and by recent numerical models of cluster formation. In the future it will be important to investigate the nature of this VLSB population further and to try and identify similar galaxies in other environments.

\end{document}